\documentclass[conference]{IEEEtran}

\usepackage{tikz}
\usepackage{color}
\usetikzlibrary{arrows, calc, snakes}
\usepackage{cite}
\usepackage{graphicx}
\usepackage[cmex10]{amsmath}
\usepackage{amsfonts}
\usepackage{amssymb}
\usepackage{algorithmic}
\usepackage{array}
\usepackage{mdwmath}
\usepackage{mdwtab}
\usepackage{eqparbox}
\usepackage{fixltx2e}
\usepackage{url}
\usepackage{enumerate}

\usepackage{dblfloatfix}
\usepackage{subfigure}
\usepackage[justification=centering]{caption}

\usepackage{relsize}

\newtheorem{definition}{Definition}
\newtheorem{theorem}{Theorem}
\newtheorem{lemma}{Lemma}

\newtheorem{protocol}{Protocol}

\newcommand{\lowerb}{\text{LB}}
\newcommand{\upperb}{\text{UB}}
\newcommand{\extra}{\text{ex}}

\begin{document}

\title{Private Data Transfer over a Broadcast Channel}

\author{\IEEEauthorblockN{Manoj Mishra, Tanmay Sharma, Bikash K. Dey}
\IEEEauthorblockA{Indian Institute of Technology Bombay, Mumbai\\
\{mmishra, tanmaysharma53, bikash\}@ee.iitb.ac.in}
\and
\IEEEauthorblockN{Vinod M. Prabhakaran}
\IEEEauthorblockA{Tata Institute of Fundamental Research, Mumbai\\
vinodmp@tifr.res.in}
}

\maketitle

\begin{abstract}

We study the following private data transfer problem: Alice has a database of files. Bob and Cathy want to access a file each from this database (which may or may not be the same file), but each of them wants to ensure that their choices of file do not get revealed even if Alice colludes with the other user.  Alice, on the other hand, wants to make sure that each of Bob and Cathy does not learn any more information from the database than the files they demand (the identities of which will be unknown to her).  Moreover, they should not learn any information about the other files even if they collude.

It turns out that it is impossible to accomplish this if Alice, Bob, and Cathy have access only to private randomness and noiseless communication links. We consider this problem when a binary erasure broadcast channel with independent erasures is available from Alice to Bob and Cathy in addition to a noiseless public discussion channel. We study the file-length-per-broadcast-channel-use rate in the honest-but-curious model. We focus on the case when the database consists of two files, and obtain the optimal rate. We then extend to the case of larger databases, and give upper and lower bounds on the optimal rate.

\end{abstract}

\IEEEpeerreviewmaketitle


\section{Introduction}
\label{sec:intro}

We consider the following problem: Alice has a database of files
(e.g., she runs a video-on-demand service). Bob and Cathy are her customers
who want to access a file each from this database, but they want to ensure that
their choices of file are not revealed, even if Alice colludes with the other
customer. Alice, on the other hand, wants to make sure that each of her
customers does not learn any more information from the database than the files
they have demanded (the identities of which will be unknown to her), and if the
customers collude they do not learn any more than the collection of files they
asked for. We will require that the privacy guarantees are unconditional (i.e.,
information theoretic). We call this the {\em private data transfer} problem.

This problem is an instance of \emph{secure multiparty computation}
(SMPC)~\cite{CramerDN}, where several mutually distrusting users wish to
communicate with each other over a network in order to compute functions of
their distributed, private inputs. At the end of such a computation, no user
learns any more information about any private data than what is revealed by its
own input and output.

It is known that for unconditionally secure computation of general
functions, private randomness and noiseless communication are insufficient
\cite{kushilevitz1992}. This holds even when the users are
honest-but-curious, i.e., they follow the protocol faithfully, but will
infer forbidden information from the random variables they accumulate over
the protocol's execution. Indeed, it can be shown that private data
transfer described above cannot be achieved if Alice, Bob, and Cathy only
have private randomness and noiseless communication (pairwise and/or
public). Additional noisy resources, in particular a noisy channel, have
been proposed~\cite{CrepKilian1988} as a resource to enable secure
computation in such settings. In this paper we will consider a (noisy) broadcast
channel from Alice to Bob and Cathy as a resource for achieving private
data transfer.

We study private data transfer over binary erasure broadcast channels for
databases of size two. There are several problems which are very closely
related to our problem.
\begin{enumerate}
\item[(i)] {\em Oblivious transfer} (OT) is a family of two-party secure
computation primitives, a specific version (namely \emph{$1$-of-$2$ string
OT}), is as follows: Alice and Bob are two-parties with Alice having $2$
equal length strings of which Bob wants exactly one string without Alice
finding out the identity of the string Bob wants. Alice wants to ensure
that Bob receives information about only one of the two strings. The
connection to our problem will be explored in greater length below.

\item[(ii)] {\em Private information retrieval} (PIR): Our problem can be
viewed as a version of the PIR problem~\cite{ChorKuGoSu98,Yekhanin10} with
symmetric privacy requirements. In the PIR problem (with asymmetric privacy
requirement), a user wants to retrieve an element from a database held by one
or more servers such that each server does not learn the identity of the
database element retrieved.  The symmetric version, where the servers also want
to ensure that the user does not learn anything more than the element
retrieved, has also been studied. The key difference with our work is that
previous works have considered only noiseless communication. Under this, it is
impossible to achieve PIR with a single server (as in our problem setting) with
an information theoretic guarantee even for the asymmetric privacy requirement.
The standard approach is to consider multiple servers (who all do not collude).
Here, we consider a single-server PIR problem with symmetric privacy
requirements in the honest-but-curious setting, but allow the use of a (noisy)
broadcast channel.

\end{enumerate}

To achieve OT, it is known that a noisy resource such as a noisy channel
between Alice and Bob is necessary, even when Alice and Bob are honest but curious. For the $1$-of-$2$ string OT described
above, OT capacity of a discrete memoryless channel (DMC) is the largest string
length (in bits) that Bob can obtain per use of the DMC. For honest-but-curious
users, Nascimento and Winter~\cite{NascWinter2008} obtained a lower bound on
the string OT capacity of DMCs and source distributions. 

Ahlswede and
Csisz{\'a}r~\cite{ot2007} obtained lower bounds on the string OT capacity of generalized
erasure channels when users are honest-but-curious. For erasure probability at least $\frac{1}{2}$, these lower bounds are tight. Pinto et.
al.~\cite{PintoDowsMorozNasc2011} showed that, for erasure probability at least $\frac{1}{2}$, the capacity of this model remains unchanged even when the parties are \emph{malicious}, that is, even when the parties may arbitrarily deviate from the protocol.

This $2$ party string OT setup was generalized to the case of a wiretapped
channel and the honest-but-curious OT capacity of the case of binary erasure
broadcast channels was characterized both for $2$-privacy (where the
eavesdropper might collude with either user) and $1$-privacy (no collusion
allowed) in~\cite{MishraDPDisit14}. A further generalization is when Alice-Bob
and Alice-Cathy want to perform {\em independent} OTs using a (noisy)
broadcast channel from Alice to Bob and Cathy, i.e.,  Alice has two pairs of
strings, Bob is necessarily interested in a file from the first pair and
Cathy from the second pair. Mishra el al~\cite{MishraDPDitw-invited14}
studied the optimal trade-off between the rates of the first pair and the
second pair for a binary erasure broadcast channel and obtained inner and outer
bounds for the $2$-privacy rate-region in the honest-but-curious setting.

Our data transfer problem can be seen as a variant of the setup
of~\cite{MishraDPDitw-invited14}, where Alice now has a collection of $N$
strings. Bob and Cathy each want to independently pick up one of the strings.
A straight forward approach for $N=2$ is to invoke the achievable scheme
of~\cite{MishraDPDitw-invited14} for the symmetric rate point by setting both
pairs as the same. However, this turns out to be sub-optimal, in general. We
propose a scheme and prove its optimality. For the general $N$ case we give 
upper and lower bounds for the optimal rate.


Section~\ref{sec:prob_defn} defines the problem for the case of a database with
two files and gives our main result which completely resolves this problem.
In section~\ref{sec:achievability}, we describe the protocol which is used to
prove the achievability part of our main result. Appendix~\ref{apndx:converse_main_result} has the proof of the converse part of our main result. The result is extended to
the case of a database with more than two files in
Section~\ref{sec:not_so_main_achievability} where we give upper and lower
bounds on the optimal rate.


\section{Problem Statement and Main Result for a Database with Two Files}
\label{sec:prob_defn}

\begin{figure}[h]
\setlength{\unitlength}{1cm}
\centering
\begin{tikzpicture}[scale=1]

\draw (1,3) rectangle (2,3.5);
\draw (3,3) rectangle (4.5,3.5);
\draw (6.5,2) rectangle (7.5,2.5);
\draw (6.5,3) rectangle (7.5,3.5);

\draw [->] (2,3.25) -- (3,3.25 );
\draw [->] (4.5,3.4) -- (6.5, 3.4);
\draw (4.5, 3.1) -- (5,3.1);
\draw [->] (5,3.1) |- (6.5,2.15);
\draw [->] (1.5,4) -- (1.5,3.5);
\draw [->] (1.5,4) -| (7,3.5);
\draw [<->] (6.1,4) |- (6.5, 2.4);

\node at (1.5,3.25) {\small{Alice}};
\node at (7,3.25) {\small{Bob}};
\node at (7,2.25) {\small{Cathy}};
\node at (3.75, 3.25) {$p_{YZ|X}$};
\node [above] at (4.25,4) {\small{Public Channel}};
\node [below] at (1.5,3) {$K_0,K_1$};
\node [below] at (7,3) {$U$};
\node [below] at (7,2) {$W$};

\node [above] at (2.5,3.25) {$X$};
\node [above] at (5.5,3.4) {$Y$};
\node [above] at (5.5,2.25) {$Z$};

\draw [->] (7.5,3.25) -- (8,3.25);
\node [right] at (8,3.25) {$\hat{K}_U$};
\draw [->] (7.5,2.25) -- (8,2.25);
\node [right] at (8,2.25) {$\hat{K}_W$};

\end{tikzpicture}
\caption{Setup for private data transfer over a broadcast channel}
\label{fig:ot-setup-bcast}
\end{figure}
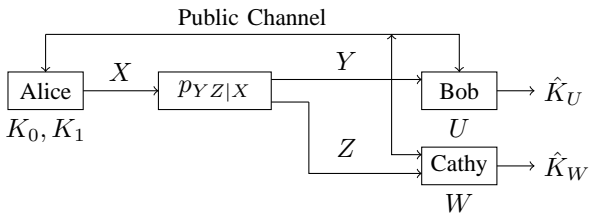

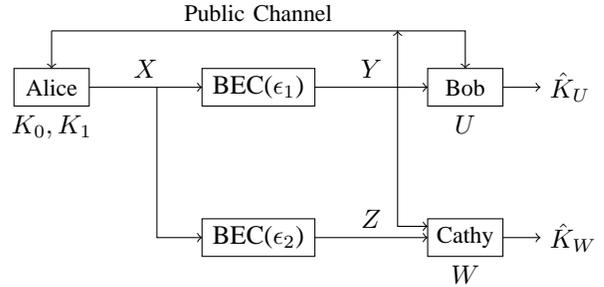
\begin{figure}[h]
\setlength{\unitlength}{1cm}
\centering
\begin{tikzpicture}[scale=1]

\draw (1,3) rectangle (2,3.5);
\draw (3.5,1) rectangle (5,1.5);
\draw (3.5,3) rectangle (5,3.5);
\draw (6.5,1) rectangle (7.5,1.5);
\draw (6.5,3) rectangle (7.5,3.5);

\draw [->] (2,3.25) -- (3.5,3.25 );
\draw [->] (5,3.25) -- (6.5, 3.25);
\draw [->] (2.9,3.25) |- (3.5,1.25);
\draw [->] (5,1.25) -- (6.5,1.25);
\draw [->] (1.5,4) -- (1.5,3.5);
\draw [->] (1.5,4) -| (7,3.5);
\draw [<->] (6.1,4) |- (6.5, 1.4);

\node at (1.5,3.25) {\small{Alice}};
\node at (7,3.25) {\small{Bob}};
\node at (7,1.25) {\small{Cathy}};
\node at (4.25, 3.25) {BEC($\epsilon_1$)};
\node at (4.25, 1.25) {BEC($\epsilon_2$)};
\node [above] at (4.25,4) {\small{Public Channel}};
\node [below] at (1.5,3) {$K_0,K_1$};
\node [below] at (7,3) {$U$};
\node [below] at (7,1) {$W$};

\node [above] at (2.75,3.25) {$X$};
\node [above] at (5.75,3.25) {$Y$};
\node [above] at (5.75,1.25) {$Z$};

\draw [->] (7.5,3.25) -- (8,3.25);
\node [right] at (8,3.25) {$\hat{K}_U$};
\draw [->] (7.5,1.25) -- (8,1.25);
\node [right] at (8,1.25) {$\hat{K}_W$};

\end{tikzpicture}
\caption{Setup for private data transfer over broadcast channel consisting of independent binary erasure channels}
\label{fig:ot-setup}
\end{figure}

For simplicity we first consider the case of a database with two files. Alice's
private database is made up of two equal sized files (bit-strings) $K_0, K_1$ which are $m$-bit
long each. 
Bob and Cathy have choice bits $U$ and $W$ respectively. $K_0, K_1,U,W$ are
independent and uniform over their respective alphabets. By $\overline{U}$ we
will denote $\overline{U}=U\oplus 1$, the complement of $U$.

The goal is for Bob to obtain $K_U$ and Cathy to obtain $K_W$ without any additional information about the database and the choice variables being revealed to any single user or pairs of users, e.g., Alice on her own should not learn anything about $U,W$; Alice and Bob working together should not learn any information about $W$; Bob on his own should not have any information about $W, K_{\overline{U}}$; Bob and Cathy working together should not learn anything about $K_{\overline{U}}$ in case $U=W$; and so on. We assume that the users are honest-but-curious.
  
In the setup in Figure~\ref{fig:ot-setup-bcast}, Alice can communicate to Bob
and Cathy over a memoryless broadcast channel $p_{Y,Z|X}$. In addition, there
is a public channel which is noiseless and has unlimited capacity. Alice, Bob
and Cathy can send messages over this public channel and each such message
will be received by all users. 

\begin{definition}
Let $n,m \in \mathbb{N}$. An $(n,m)$-\emph{protocol} is an exchange of messages between Alice, Bob, and Cathy over the setup of Figure~\ref{fig:ot-setup-bcast}. Here $m$ is the length of each bit string in Alice's private database and $n$ is the number of uses of the broadcast channel she makes. Before each channel transmission and also after the last channel transmission, Alice, Bob and Cathy can exchange an arbitrary but finite (with probability 1) number of messages over the public channel, taking turns to send each such message. The messages exchanged over the public channel and the channel transmissions are allowed to be randomized, but the parties may only use private randomness to accomplish this. The \emph{rate} $R$ of an $(n,m)$-protocol is defined to be $R  := {m}/{n}$.
\end{definition}
We denote by $\mathbf{F}$ the transcript of the public channel at the end of an ($n,m$)-protocol.

\begin{definition}
The \emph{final view} of a user is the set of random variables that the user observes or generates over the duration of the $(n,m)$-protocol. The final views of Alice, Bob and Cathy are, respectively,
\begin{align}
V_A & := (K_0, K_1,X^n,\mathbf{F}), \label{eq:VA}\\
V_B & := (U,Y^n,\mathbf{F}), \text{ and} \label{eq:VB}\\
V_C & := (W,Z^n,\mathbf{F}). \label{eq:VC}
\end{align}
\end{definition}

\begin{definition}
A rate $R$ is an \emph{achievable $2$-private data transfer rate} if there exists a sequence of $(n,m)$-protocols with rate $R$ such that as $n \longrightarrow \infty$, we have 
\begin{align}
 P[\hat{K}_U \neq K_U \text{ or } \hat{K}_W \neq K_W] & \longrightarrow 0 \label{eqn:ach_rate_1}\\
 I(K_{\overline{U}}; V_B,V_C| U = W ) & \longrightarrow 0 \label{eqn:ach_rate_3}\\
 I(U ; V_A,V_C) & \longrightarrow 0 \label{eqn:ach_rate_4} \\
 I(W ; V_A,V_B) & \longrightarrow 0 \label{eqn:ach_rate_5} \\
 I(U,W; V_A) & \longrightarrow 0 \label{eqn:ach_rate_6} \\
 I(W,K_{\overline{U}} ; V_B) & \longrightarrow 0 \label{eqn:ach_rate_7}\\
 I(U,K_{\overline{W}} ; V_C) & \longrightarrow 0. \label{eqn:ach_rate_8}
\end{align}
\end{definition}

\begin{definition}
The \emph{$2$-private data transfer capacity} $C_{2P}$ for the setup of Figure~\ref{fig:ot-setup-bcast} is the supremum of all achievable $2$-private data transfer rates.
\end{definition}
In this paper, we study the specific instance of independent binary erasure broadcast channel (shown in Figure~\ref{fig:ot-setup}), where $p_{YZ|X} = p_{Y|X} \cdot p_{Z|X}$ and where $p_{Y|X}$ is a binary erasure channel BEC($\epsilon_1$) with erasure probability $\epsilon_1$, and $p_{Z|X}$ is a BEC($\epsilon_2$).



Our main result is a characterization of the 2-private data transfer capacity of the independent erasure broadcast channel.
\begin{theorem} \label{thm:main}
\begin{equation*}
C_{2P} = \min \left(  \epsilon_2  (1 - \epsilon_1 ), \;  \epsilon_1 (1 - \epsilon_2 ), \; \epsilon_1 \epsilon_2 \right).
\end{equation*}
\end{theorem}
We prove this theorem in the next section by giving a protocol which can achieve rates arbitrarily close to capacity and proving a converse.


\section{Proof of Theorem~\ref{thm:main}}
\label{sec:achievability}

In this section, we first describe a protocol which will be used to achieve
$2$-private data transfer capacity of the setup of
Figure~\ref{fig:ot-setup}. We note that the protocol described for the setup in \cite{MishraDPDitw-invited14}, though useful for the private data transfer problem here, does not (in general) achieve the $2$-private data transfer capacity of the setup of Figure~\ref{fig:ot-setup} (eg. consider $\epsilon_1 < \frac{1}{2}, \epsilon_2 \in (\frac{1}{2}, \frac{2}{3})$). Before giving a formal description of our
protocol, we will outline its main ideas.

Alice begins by transmitting a sequence $X^n$ of independent, uniformly
distributed bits, indexed by $1,2,\ldots,n$, over the broadcast channel.
Bob and Cathy receive independently erased versions $Y^n$ and $Z^n$,
respectively, of the transmitted bits. 

Let us consider the case $\epsilon_1,\epsilon_2 \leq 1/2$.
Bob has about $n\epsilon_1$ erased bits in $Y^n$, and he takes the {\em
indices} of these bits as the \emph{bad set} $B$. Out of the indices of
unerased bits in $Y^n$, Bob randomly picks a subset of indices, of the same
{cardinality} as $B$, and calls it the \emph{good set} $G$. If $U = 0$, Bob assigns
($L_0, L_1$) = ($G,B$), otherwise Bob assigns ($L_0, L_1$) = ($B,G$). Bob
sends ($L_0, L_1$) over the public channel. Notice that even if Alice and
Cathy get together, they will not learn $U$ from ($L_0,L_1$) that Bob
sent over the public channel. This follows from the independence of the
erasure channels to Bob and Cathy and the memoryless nature of
erasures.

Cathy confines her attention to $Z^n|_{L_0 \cup L_1}$, the restriction of
$Z^n$ to the indices in $L_0 \cup L_1$.  In a manner similar to Bob, out of
$Z^n|_{L_0 \cup L_1}$, Cathy forms her own good and bad sets of indices
$\tilde{G}, \tilde{B}$ {respectively}, each of size about $2n\epsilon_1\epsilon_2$. If $W
= 0$, Cathy assigns ($\tilde{L}_0, \tilde{L}_1$) =
($\tilde{G},\tilde{B}$), otherwise Cathy assigns ($\tilde{L}_0,
\tilde{L}_1$) = ($\tilde{B},\tilde{G}$). Cathy sends ($\tilde{L}_0,
\tilde{L}_1$) over the public channel. 

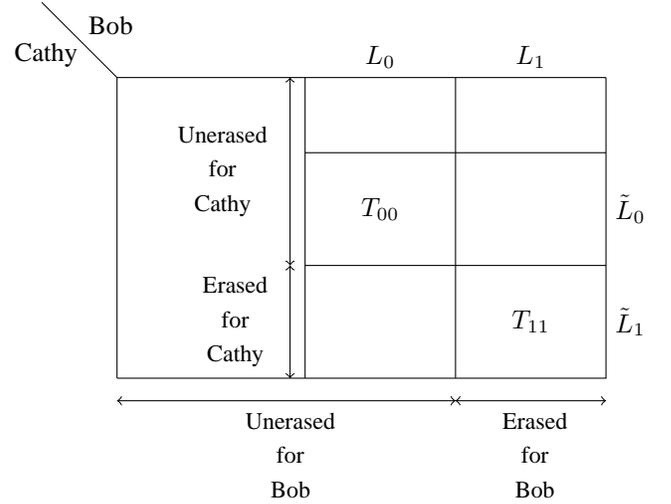
\begin{figure}[h]
\setlength{\unitlength}{1cm}
\centering
\begin{tikzpicture}[scale=1]

\draw (1,1) |- (7.5,5);
\draw (7.5,5) |- (1,1);
\draw (0,6) -- (1,5);

\draw (5.5,5) -- (5.5,1);
\draw (3.5,5) -- (3.5,1);
\draw (3.5,4) -- (7.5,4);
\draw (3.5,2.5) -- (7.5,2.5);

\node [right] at (0.5,5.7) {Bob};
\node [left] at (0.6,5.3) {Cathy};

\node [above] at (4.5,5) {$L_0$};
\node [above] at (6.5,5) {$L_1$};
\node[right] at (7.5, 3.25) {$\tilde{L}_0$};
\node [right] at (7.5, 1.75) {$\tilde{L}_1$};

\node at (4.5, 3.25) {$T_{00}$};
\node at (6.5, 1.75) {$T_{11}$};

\draw [<->] (1,0.7) -- (5.5,0.7);
\draw [<->] (5.5,0.7) -- (7.5,0.7); 
\node [below] at (3.25,0.7)  {\small{ $\begin{array}{c} \text{Unerased} \\ \text{for} \\ \text{Bob} \end{array}$ }};
\node [below] at (6.5,0.7) {\small{ $\begin{array}{c} \text{Erased} \\ \text{for} \\ \text{Bob} \end{array}$ }};

\draw [<->] (3.3,5) -- (3.3,2.5);
\draw [<->] (3.3,2.5) -- (3.3, 1);
\node [left] at (3.3,3.75) {\small{ $\begin{array}{c} \text{Unerased} \\ \text{for} \\ \text{Cathy} \end{array}$ }};
\node [left] at (3.3, 1.75) {\small{ $\begin{array}{c} \text{Erased} \\ \text{for} \\ \text{Cathy} \end{array}$ }};

\end{tikzpicture}
\caption{Illustration of the sets used in the protocol when $U=W=0$ and $\epsilon_1, \epsilon_2 \leq \frac{1}{2}$}
\label{fig:ot_keys_smallfig}
\end{figure}

Alice forms two \emph{data transfer (DT) keys} $T_{00}$ and  $T_{11}$ as (also see Figure~\ref{fig:ot_keys_smallfig}):
\begin{subequations}
\begin{align}
T_{00} & = X^n|_{L_0\cap \tilde{L}_0} \\ 
T_{11} & = X^n|_{L_1 \cap \tilde{L}_1}
\end{align} 
\label{eq:T}
\end{subequations}

Alice then sends the following encrypted strings over the public channel :
      \begin{align*}
        M_0 = K_0 & \oplus T_{00}, \\ 
        M_1 = K_1 & \oplus T_{11}.
      \end{align*}
Bob knows $T_{UU}$. Hence, using $M_U$, Bob can recover $K_U$.
Also, Cathy knows $T_{WW}$. Hence, using $M_W$, Cathy can
recover $K_W$. Bob, however, does not know anything about
$T_{\overline{U}\overline{U}}$, and since $K_{\overline{U}}$
is encrypted with $T_{\overline{U}\overline{U}}$,
he does not learn anything about $K_{\overline{U}}$. Similarly,
Cathy does not learn anything about $K_{\overline{W}}$. If $U=W$, then
even if Bob and Cathy get together, they cannot learn anything about
$K_{\overline{U}}$ since $T_{\overline{U}\overline{U}}$ is erased for both
of them.

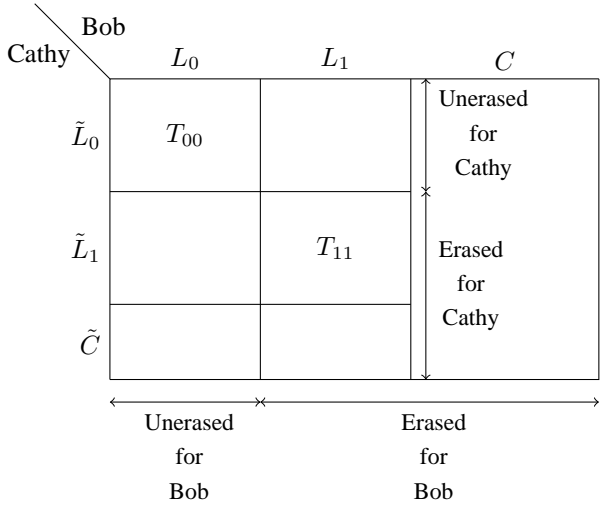
\begin{figure}[h]
\setlength{\unitlength}{1cm}
\centering
\begin{tikzpicture}[scale=1]

\draw (1,1) |- (7.5,5);
\draw (7.5,5) |- (1,1);
\draw (0,6) -- (1,5);
\draw (3,5) -- (3,1);
\draw (5,5) -- (5,1);
\draw (1,3.5) -- (5,3.5);
\draw (1,2) -- (5,2);

\node [right] at (0.5,5.7) {Bob};
\node [left] at (0.6,5.3) {Cathy};

\node [above] at (2,5) {$L_0$};
\node [above] at (4,5) {$L_1$};
\node [above] at (6.25,5) {$C$};
\node [left] at (1, 4.25) {$\tilde{L}_0$};
\node [left] at (1, 2.75) {$\tilde{L}_1$};
\node [left] at (1, 1.5) {$\tilde{C}$};

\node at (2, 4.25) {$T_{00}$};
\node at (4, 2.75) {$T_{11}$};

\draw [<->] (1,0.7) -- (3,0.7);
\draw [<->] (3,0.7) -- (7.5,0.7); 
\node [below] at (2,0.7)  {\small{ $\begin{array}{c} \text{Unerased} \\ \text{for} \\ \text{Bob} \end{array}$ }};
\node [below] at (5.25,0.7) {\small{ $\begin{array}{c} \text{Erased} \\ \text{for} \\ \text{Bob} \end{array}$ }};

\draw [<->] (5.2,5) -- (5.2,3.5);
\draw [<->] (5.2,3.5) -- (5.2, 1);
\node [right] at (5.2,4.25) {\small{ \hspace{-10pt} $\begin{array}{c} \text{Unerased} \\ \text{for} \\ \text{Cathy} \end{array}$ }};
\node [right] at (5.2, 2.25) {\small{ \hspace{-10pt} $\begin{array}{c} \text{Erased} \\ \text{for} \\ \text{Cathy} \end{array}$ }};

\end{tikzpicture}
\caption{Illustration of the sets used in the protocol when $U=W=0$ and $\epsilon_1, \epsilon_2 > \frac{1}{2}$}
\label{fig:ot_keys}
\end{figure}

When $\epsilon_1, \epsilon_2 > \frac{1}{2}$, the size of $L_0,L_1$ is 
about $n(1-\epsilon_1)$ each, and the size of $\tilde{L}_0,\tilde{L}_1$ is
about $2n(1-\epsilon_2)(1-\epsilon_1)$ each.
Bob and Cathy have
additional erased indices that they did not use for sets $B$ and $\tilde{B}$
respectively. Bob forms the {set} $C$ (of size $n(2\epsilon_1 - 1)$) and
Cathy forms {the set} $\tilde{C}$ (of size about $2{n}(1 - \epsilon_1)(2\epsilon_2 -
1)$) out of these unused erased indices (see Figure~\ref{fig:ot_keys}) and
declare them over the public channel. Thereafter, Alice-Bob get an additional
rate using a two-party oblivious transfer (OT) protocol~\cite{ot2007} over
$X^n|_{\tilde{C}}$.  Notice that a two-party protocol is appropriate since
bits in $\tilde{C}$ are guaranteed to be erased for Cathy. Similarly,
Alice-Cathy get additional rate using a two-party OT protocol over
$X^n|_C$. Thus, for $\epsilon_1, \epsilon_2 > \frac{1}{2}$, the protocol
will rate-split the string $K_0$ as $(\dot{K}_0,\ddot{K}_0)$ (and similarly
for $K_1$) of appropriate lengths to perform the data transfer in two
parts. However, for all other regimes of $\epsilon_1, \epsilon_2$,
$\dot{K}_0 = K_0$ and $\dot{K}_1 = K_1$.

We now give a step-wise description of the protocol. See Appendix~\ref{apndx:sizes_and_extra_rate} for more details on the set sizes and rate calculations mentioned in this protocol.
\begin{protocol}

\label{sch:our_2P}

Let $\delta > 0$. Let $r_1 = \min \{\epsilon_1, 1 - \epsilon_1\} - \delta$ and $r_2 = \min \{\epsilon_2, 1 - \epsilon_2\} - \delta$.

\begin{description}
\item[\textbf{Alice}] Transmits a sequence $X^n$ of independent, uniformly
distributed bits over the broadcast channel.

\item[\textbf{Bob}] Receives $Y^n$ from BEC($\epsilon_1$). Bob's set of
erased and unerased indices are
       \begin{align*}
          E  &:= \{ i \in \{1, 2, \ldots, n\} : Y_i = \text{erasure} \},\\
         \overline{E}  &:= \{ i \in \{1, 2, \ldots, n\} : Y_i \neq \text{erasure} \}.
       \end{align*}
If $|E|<n(\epsilon_1-\delta)$ or $|\overline{E}|<n(1-\epsilon_1-\delta)$, Bob declares error.
Otherwise Bob randomly picks the following sets:
              \begin{IEEEeqnarray*}{rCl}
               G & \thicksim & \text{Unif} \left \{ A \subseteq \overline{E} : |A| = nr_1 \right \} , \\
               B & \thicksim &  \text{Unif} \left \{ A \subseteq E : |A| = nr_1 \right \} .\\
               \text{If $\epsilon_1,\epsilon_2 > \frac{1}{2}$} & & \\
                  C & \thicksim & \text{Unif} \left \{ A \subseteq (E \backslash B) : |A| = n(2\epsilon_1 - 1) \right \} \\
               \text{else} && \\
                  C & = & \emptyset.
              \end{IEEEeqnarray*}
      
     Now, depending on the value of $U$, Bob further creates the sets $L_0, L_1$ as follows.
	      \begin{align*}
	       U = 0:\qquad &L_0 = G,\quad L_1 = B \\
               U = 1:\qquad &L_0 = B,\quad L_1 = G
        \end{align*}
     Bob sends $L_0,L_1,C$ over the public channel.

\item [\textbf{Cathy}] Over the subset $Z^n|_{L_0 \cup L_1}$, Cathy
defines her set of erased and unerased indices as
    \begin{align*}
      E' &:= \{ i \in L_0 \cup L_1 : Z_i = \text{erasure} \} \\
      \overline{E}' &:= \{ i \in L_0 \cup L_1 : Z_i \neq \text{erasure} \}
    \end{align*}
    {If $|E'|< 2nr_1(\epsilon_2 - \delta)$ or $|\overline{E}'|< 2nr_1(1 - \epsilon_2 - \delta)$,
then Cathy declares error.}

    Otherwise Cathy randomly picks the following sets:
     \begin{IEEEeqnarray*}{rCl}
               \tilde{G} & \thicksim & \text{Unif} \left\{ A \subseteq \overline{E}' : |A| = 2nr_1r_2 \right\} \\
               \tilde{B} & \thicksim & \text{Unif} \left\{ A \subseteq E' : |A| = 2nr_1r_2 \right\} \\
              \text{If $\epsilon_1,\epsilon_2 > \frac{1}{2}$} & & \\
                  \tilde{C} & \thicksim & \text{Unif} \left\{ A \subseteq (E' \backslash \tilde{B}) : |A| = 2nr_1(2\epsilon_2 - 1)\right\} \\
               \text{else} && \\
                  \tilde{C} & = & \emptyset.  
      \end{IEEEeqnarray*} 
     Now, depending on the value of $W$, Cathy further creates the sets
$\tilde{L}_0, \tilde{L}_1$ as follows:
	\begin{align*}
	 W = 0:\qquad  & \tilde{L}_0 = \tilde{G}, \quad \tilde{L}_1 = \tilde{B} \\
         W = 1:\qquad & \tilde{L}_0 = \tilde{B}, \quad \tilde{L}_1 = \tilde{G}
        \end{align*}
     Cathy sends $\tilde{L}_0, \tilde{L}_1, \tilde{C}$ over the public channel.

\item [\textbf{Alice}] forms the data transfer keys $T_{00}, T_{11}$ as
in \eqref{eq:T}, and sends {the} following strings over the public channel.
      \begin{align*}
        M_0 = \dot{K}_0 & \oplus T_{00} , \\
        M_1 = \dot{K}_1 & \oplus T_{11}. \\
      \end{align*}
\item [\textbf{Bob}] knows $T_{UU}$ and, thus, can recover $\dot{K}_U$.
\item [\textbf{Cathy}] knows $T_{WW}$ and, thus, can recover $\dot{K}_W$.

\item [\textbf{Bob}] For $\epsilon_1,\epsilon_2 > \frac{1}{2}$, Bob selects
a set $\tilde{S} \subseteq \tilde{C}$ as follows: if $\epsilon_1 <
\epsilon_2$, Bob sets $\tilde{S}$ as the first $\frac{n(2\epsilon_1 - 1)r_2}{(\frac{1}{2} - \delta)}$ bits of $\tilde{C}$, otherwise Bob sets $\tilde{S} = \tilde{C}$. See Appendix~\ref{apndx:extra_rate} for more details.

Alice and Bob then follow the $2$-party OT protocol \cite{ot2007} using $X^n|_{\tilde{S}}$, with the inputs ($\ddot{K}_0, \ddot{K}_1, U$).

\item [\textbf{Cathy}] For $\epsilon_1,\epsilon_2 > \frac{1}{2}$, Cathy
selects a set $S \subseteq C$ as follows: If $\epsilon_2 < \epsilon_1$,
Cathy {sets} $S$ as the first $\frac{2nr_1(2\epsilon_2 - 1)(\frac{1}{2} - \delta)}{r_2}$ bits of $C$, otherwise Cathy {sets} $S = C$. See Appendix~\ref{apndx:extra_rate} for more details.

Alice and Cathy then follow the $2$-party OT protocol \cite{ot2007} using $X^n|_S$, with the inputs ($\ddot{K}_0, \ddot{K}_1, W$).
\end{description}

\end{protocol}
Using this protocol we obtain the following achievability result.
\begin{lemma}
\label{lem:achievable_rate}
For the setup of Figure~\ref{fig:ot-setup}, if $R < \min \left(  \epsilon_2  (1 - \epsilon_1 ), \;  \epsilon_1 (1 - \epsilon_2 ), \; \epsilon_1 \epsilon_2 \right)$, then $R$ is an achievable $2$-private data transfer rate.
\end{lemma}
The proof of this lemma is deferred to Appendix~\ref{apndx:proof_of_lemma}. The main ideas used in the proof are the following:
\begin{itemize}
\item First, by Chernoff bound, the probability that the algorithm will abort
due to the size conditions not being met is exponentially small.
\item Bob  knows $T_{UU}$. Thus, from  $\dot{K}_U \oplus T_{UU}$ Bob can recover $\dot{K}_U$.
\item Cathy knows $T_{WW}$. Thus, from  $\dot{K}_W \oplus T_{WW}$, Cathy can recover $\dot{K}_W$. 
\item When $U=W$,  colluding Bob and Cathy know nothing about $T_{\overline{U} \overline{U}}$ since it is erased for both of them. Since Alice's transmissions always encrypt  $\dot{K}_{\overline{U}}$ with $T_{\overline{U} \overline{U}}$, colluding Bob and Cathy learn nothing about $\dot{K}_{\overline{U}}$.
\item Alice never learns either $U$ or $W$. Note that Alice can learn $U$ or $W$ only from the sets of indices she receives from Bob and Cathy. In the setup, the channels act independently of each other and independently on each input bit.  Further, the protocol ensures $|L_0| = |L_1|$ and $|\tilde{L}_0| = |\tilde{L}_1|$. Thus, Alice has no means of learning about which sets of indices it receives correspond to erasures. Also, since Alice learns nothing about $U$, we can show that colluding Alice and Cathy cannot learn anything about $U$ either. Similarly, since Alice learns nothing about $W$, colluding Alice and Bob cannot learn anything about $W$.
\end{itemize}

\subsection*{Converse of Theorem~\ref{thm:main}}
The converse of Theorem~\ref{thm:main} is proved in Appendix~\ref{apndx:converse_main_result}, where we show the following general upper bound on $C_{2P}$ in the setup of Figure~\ref{fig:ot-setup-bcast}:
\begin{align*}
 &C_{2P} \leq \nonumber \\
 &\min\left( \max_{p_X} I(X ; Y | Z),\max_{p_X} I(X ; Z | Y), \max_{p_X} H(X | Y, Z) \right).
\end{align*}
Evaluated for the setup of Figure~\ref{fig:ot-setup}, this gives the required upper bound.


\section{Databases with $N>2$ files}
\label{sec:not_so_main_achievability}

The problem definition in Section~\ref{sec:prob_defn} can be readily extended to a database with $N$ files; see Appendix~\ref{sec:general:setup}. Generalizing the protocol and the converse (see Appendix~\ref{apndx:not_so_main_achievability}) from the last section we can obtain the following upper and lower bounds on the 2-private data transfer capacity.
Let
\begin{align*}
R_{\upperb} =
\min\left(\epsilon_2(1-\epsilon_1),\; \epsilon_1 (1 - \epsilon_2),\; \frac{\epsilon_1 \epsilon_2}{N-1} \right),
\end{align*}%
and
\begin{align*}
R_{\lowerb} = \left\{  \begin{array}{l l} \frac{\epsilon_1}{N-1} \cdot \frac{\epsilon_2}{N-1}, &  \epsilon_1,\epsilon_2 \leq \frac{N-1}{N} \\ 
                                                       \frac{\epsilon_1}{N-1} \cdot (1 - \epsilon_2), & \epsilon_1 \leq \frac{N-1}{N}, \epsilon_2 > \frac{N-1}{N} \\
                                                       \frac{\epsilon_2}{N-1} \cdot (1 - \epsilon_1), & \epsilon_1 > \frac{N-1}{N}, \epsilon_2 \leq \frac{N-1}{N} \\
                                                       (1 - \epsilon_1) \cdot (1 - \epsilon_2) + R_{\extra}, & \epsilon_1,\epsilon_2 > \frac{N-1}{N}, 
                              \end{array}
            \right.
\end{align*}
where 
\begin{align*}
      R_{\extra} = \min( (1-\epsilon_2)(1 - N(1-\epsilon_1)), (1-\epsilon_1) (1 - N(1-\epsilon_2)) ).
\end{align*}

\begin{theorem}
\label{thm:not_so_main}
\begin{equation*}
R_{\lowerb} \leq C_{2P} \leq R_{\upperb}.
\end{equation*}
\end{theorem}

We note that the upper and lower bounds in Theorem~\ref{thm:not_so_main} are not very close, especially for large $N$. For instance, for erasure probabilities less that $1-\frac{1}{N}$, there is a factor of $(N-1)$ gap.


\section{Future Work}

Besides finding tighter bounds for the general $N$ case, there are several
natural directions of enquiry: (i) the case of more than two users, (ii)
asymmetric case where privacy is desired only on the choices, (iii) other
channel models, (iv) the malicious model where the dishonest users may deviate
from the protocol arbitrarily. 

\section{Acknowledgements}
The work  was supported in part by the Bharti Centre for Communication, IIT Bombay, a grant from the Department of Science and Technology, Government of India, to IIT Bombay, and by Information Technology Research Academy (ITRA), Government of India under ITRA-Mobile grant ITRA/15(64)/Mobile/USEAADWN/01.  V. Prabhakaran's research was also supported in part by a Ramanujan Fellowship from the Department of Science and Technology, Government of India.



\appendices

\section{Proof of Lemma~\ref{lem:achievable_rate}}
\label{apndx:proof_of_lemma}

In this proof, we use a sequence $\{ \mathcal{P}_n \}_{n \in \mathbb{N}}$ of Protocol~\ref{sch:our_2P} and show that (\ref{eqn:ach_rate_1}) - (\ref{eqn:ach_rate_8}) hold for $\{ \mathcal{P}_n \}_{n \in \mathbb{N}}$. We consider the case when either $\epsilon_1 \leq \frac{1}{2}$ or $\epsilon_2 \leq \frac{1}{2}$. The case where both $\epsilon_1, \epsilon_2 > \frac{1}{2}$ involves an additional phase (as described in Section~\ref{sec:achievability}) where the well-understood $2$-party OT protocol of \cite{ot2007} is invoked. For ease of exposition, this case is not being considered here. Hence, for the proof presented here, $\dot{K}_0 = K_0$ and $\dot{K}_1 = K_1$.

For the protocol $\mathcal{P}_n$, we get $r_n = r_1r_2 \longrightarrow C_{2P}$, since $\delta > 0$ can be chosen arbitrarily small for sufficiently large $n$. 

Let $J$ denote the event that either Bob or Cathy declares an error during the protocol. Then, by Chernoff bound, $P[J=1] \longrightarrow 1$ as $n \longrightarrow \infty$.

\begin{enumerate}

\item To show that (\ref{eqn:ach_rate_1}) is satisfied for $\{P_n\}_{n \in \mathbb{N}}$, we note that
\begin{align*}
P[\hat{K}_U & \neq K_U \text{ or } \hat{K}_W \neq K_W] \\ & = P[J=0]P[ \hat{K}_U \neq K_U \text{ or } \hat{K}_W \neq K_W | J = 0] \\ & \quad + P[J=1]P[ \hat{K}_U \neq K_U \text{ or } \hat{K}_W \neq K_W | J = 1]
\end{align*} 

Since $Pr[J=0] \rightarrow 0$ exponentially fast, it is sufficient to show that $P[ \hat{K}_U \neq K_U \text{ or } \hat{K}_W \neq K_W | J = 1] \longrightarrow 0$ as $n \longrightarrow \infty$.

Now, when $J=1$, Bob knows $T_{UU}$ and, thus, recovers $\dot{K}_U$. Similarly, Cathy knows $T_{WW}$ and, thus, recovers $\dot{K}_W$. As a result, $P[ \hat{K}_U \neq K_U \text{ or } \hat{K}_W \neq K_W | J = 1] = 0$.

\end{enumerate}

For the remaining part of this proof, we define the following quantities for ease of notation:

\begin{IEEEeqnarray*}{rCl}
\dot{\mathbf{G}} & = & (G, B, \tilde{G}, \tilde{B}),\\
\dot{\mathbf{L}} & = & (L_0,L_1,\tilde{L}_0, \tilde{L}_1) \\
\dot{\mathbf{M}} & = & (M_0, M_1) \\
\dot{\mathbf{F}} & = & ( \dot{\mathbf{L}},  \dot{\mathbf{M}})
\end{IEEEeqnarray*}

\begin{enumerate}
\setcounter{enumi}{1}

\item  To show that (\ref{eqn:ach_rate_3}) is satisfied for $\{P_n\}_{n \in \mathbb{N}}$, we note that

\begin{align*}
I(K_{\overline{U}} & ; V_B,V_C|U=W ) \\ 
                           & \leq I(K_{\overline{U}} ; V_B,V_C, J|U=W) \\
                           & = \sum_{j=0,1}Pr[J=j] \,I(K_{\overline{U}} ; V_B,V_C| J=j,U=W) \\
                           & \quad + I(K_{\overline{U}} ; J|U=W) .
\end{align*}
Since $Pr[J=0] \rightarrow 0$ exponentially fast and $I(K_{\overline{U}} ; J|U=W)=0$, it is sufficient to show that $I(K_{\overline{U}} ; V_B,V_C| U=W, J=1)) \longrightarrow 0$ as $n \longrightarrow \infty$. Now,

\begin{align*}
& I(K_{\overline{U}}; V_B, V_C| U = W, J = 1)\\
& =  H(K_{\overline{U}}|U =W, J = 1)-H(K_{\overline{U}}|V_B, V_C,U =W, J = 1)\\
& =  H(K_{\overline{U}})-H(K_{\overline{U}}|V_B, V_C,U =W, J = 1)\\
& =  H(K_{\overline{U}})-H(\dot{K}_{\overline{U}}|V_B, V_C,U =W, J = 1)\\
& =  H(K_{\overline{U}})-H(\dot{K}_{\overline{U}}| U,W,Y^n,Z^n, \dot{\mathbf{F}} ,U =W, J = 1)\\
& =  H(K_{\overline{U}})-H(\dot{K}_{\overline{U}}| U,W,Y^n,Z^n, \dot{\mathbf{L}}, \dot{\mathbf{M}} ,U =W, J = 1)
\end{align*}

Now,

\begin{align*}
& H(\dot{K}_{\overline{U}} | U,W,Y^n,Z^n,  \dot{\mathbf{L}},  \dot{\mathbf{M}} , U =W,J=1)\\
& =  H(\dot{K}_{\overline{U}} | U,W,Y^n,Z^n, \dot{\mathbf{G}}, \dot{\mathbf{M}}, U =W, J = 1) \\
& =  H(\dot{K}_{\overline{U}} | U,W,Y^n,Z^n, \dot{\mathbf{G}}, \dot{\mathbf{M}}, T_{UU}, U =W,J=1)\\
& [T_{UU} \text{ is a function of }  \text{(}\dot{\mathbf{G}}, Y^n,Z^n\text{)}]  \\
& =  H(\dot{K}_{\overline{U}} | U, \dot{\mathbf{M}}, T_{UU}, U=W, J=1)\\
& \text{[ since $\dot{K}_{\overline{U}} - (U, \dot{\mathbf{M}}, T_{UU}, U=W, J=1) - (W,Y^n,Z^n,\dot{\mathbf{G}})$} \\ & \text{ is a markov chain]} \\
& =   H(\dot{K}_{\overline{U}} | U, \dot{K}_U, \dot{K}_{\overline{U}} \oplus T_{\overline{U} \overline{U}}, T_{UU}, U=W, J=1) \\
& =  H(\dot{K}_{\overline{U}} | \dot{K}_{\overline{U}} \oplus T_{\overline{U} \overline{U}}) \\
& =  H(\dot{K}_{\overline{U}})
\end{align*}

So we get

\begin{align*}
I(K_{\overline{U}}; V_B, V_C| U = W, J = 1) & =  H(K_{\overline{U}}) -  H(\dot{K}_{\overline{U}}) \\
                                                                & =  0
\end{align*}


\item   To show that (\ref{eqn:ach_rate_4}) is satisfied for $\{P_n\}_{n \in \mathbb{N}}$, as before, it will suffice to show that $I( U ; V_A,V_C| J = 1) \longrightarrow 0$.

\begin{align*}
& I(U  ;  V_A,V_C | J = 1) \\
 & =  I(U ; K_0,K_1,W,X^n,Z^n,\dot{\mathbf{F}} | J = 1) \\
 & =  I(U ; K_0,K_1,W,X^n,Z^n, \dot{\mathbf{L}}, \dot{\mathbf{M}} | J = 1) \\
 & =  I(U ; K_0,K_1,W,X^n,Z^n, \dot{\mathbf{L}} | J = 1) \\
 & \text{[ $\dot{\mathbf{M}}$ is a function of ($K_0,K_1,X^n, \dot{\mathbf{L}}$) ]} \\
 & =  I(U ; X^n,Z^n, \dot{\mathbf{L}} | J = 1) \\
 & \text{[$ U - (X^n,Z^n,\dot{\mathbf{L}}, J = 1) - (K_0,K_1,W) $]} \\
 & =  I(U ; X^n,L_0,L_1 | J = 1) \\
 & \text{[$ U - (X^n,L_0,L_1, J = 1) - (Z^n,\tilde{L}_0, \tilde{L}_1) $]} \\
 & =  I(U ; L_0,L_1 | J = 1) \\
 & \text{[$ U - (L_0,L_1, J = 1) - X^n $]} \\
 & =  H(L_0,L_1 | J = 1) - H(L_0,L_1|U, J = 1) \\
 & =  H(L_0,L_1 | J = 1) - H(G,B|U, J = 1) \\
 & =  H(L_0,L_1 | J = 1) - H(G,B | J = 1) \\
 & =  0 \\
 & \text{[ since ($L_0,L_1$), ($G,B$) have same distribution,} \\ & \text{   conditioned on $J = 1$ ]}
\end{align*}

\item The proof for showing that (\ref{eqn:ach_rate_5}) is satisfied for $\{P_n\}_{n \in \mathbb{N}}$ is similar to showing that (\ref{eqn:ach_rate_4}) is satisfied for $\{P_n\}_{n \in \mathbb{N}}$.

\item   To show that (\ref{eqn:ach_rate_6}) is satisfied for $\{P_n\}_{n \in \mathbb{N}}$, it will suffice to show that $I(U,W ; V_A | J = 1) \longrightarrow 0$.

\begin{align*}
& I(U,W  ;  V_A | J = 1) \\
 & =  I(U,W ; K_0,K_1, X^n, \dot{\mathbf{F}} | J = 1) \\
 & =  I(U,W ; K_0,K_1, X^n, \dot{\mathbf{L}}, \dot{\mathbf{M}} | J = 1) \\
 & =  I(U,W ; K_0,K_1, X^n, \dot{\mathbf{L}} |  J = 1) \\
 & \text{[ $\dot{\mathbf{M}}$ is a function of ($K_0,K_1,X^n, \dot{\mathbf{L}}$) ]} \\
 & =  I(U,W ; \dot{\mathbf{L}} | J = 1) \\
 & \text{[$ U,W - (\dot{\mathbf{L}}, J = 1) - (K_0,K_1,X^n) $]} \\
 & =  I(U ; L_0,L_1 | J = 1) + I(W ; \tilde{L}_0, \tilde{L}_1 | J = 1) \\
 & =  H(L_0,L_1 | J = 1) - H(G,B | J = 1) \\ & \quad + H(\tilde{L}_0, \tilde{L}_1 | J = 1) - H(\tilde{G}, \tilde{B} | J = 1) \\
 & =  0 \\
 & \text{[ since ($L_0,L_1$), ($G,B$) have same distribution } \\
 & \text{ and ($\tilde{L}_0,\tilde{L}_1$), ($\tilde{G},\tilde{B}$) have same distribution} \\ & \text{  conditioned on $J = 1$ ]}
\end{align*}

\item    To show that (\ref{eqn:ach_rate_7}) is satisfied for $\{P_n\}_{n \in \mathbb{N}}$, it will suffice to show that $I(W , K_{\overline{U}} ; V_B | J = 1) \longrightarrow 0$.

\begin{align*}
& I(W  ,  K_{\overline{U}} ; V_B | J = 1) \\
 & =  I(W, K_{\overline{U}} ; V_B, T_{UU} | J = 1) \\
 & \text{[ since $T_{UU}$ is a function of $V_B$ ]} \\
 & =  I(W, K_{\overline{U}} ; V_B, K_U, T_{UU} | J = 1) \\
 & \text{[ since $K_U$ is a function of ($V_B, T_{UU}$) ]} \\
& =  I(W, K_{\overline{U}} ; U, Y^n, \dot{\mathbf{L}},  \dot{\mathbf{M}},  K_U, T_{UU} | J = 1) \\
 & =  I(W, K_{\overline{U}} ; U, Y^n, \dot{\mathbf{L}}, K_U, T_{UU}, K_{\overline{U}} \oplus T_{\overline{U}\overline{U}} | J = 1) \\
 & =  I(W, K_{\overline{U}} ; \tilde{L}_0, \tilde{L}_1, K_{\overline{U}} \oplus T_{\overline{U}\overline{U}} | J = 1) \\
& \text{ [ $(W, K_{\overline{U}}) - (\tilde{L}_0, \tilde{L}_1,K_{\overline{U}} \oplus T_{\overline{U}\overline{U}}, J=1)$} \\ & \quad \text{$- (U,Y^n,L_0,L_1,K_U, T_{UU})$ ] } \\
 & =  I(W ; \tilde{L}_0, \tilde{L}_1 | J = 1) + I(K_{\overline{U}} ; K_{\overline{U}} \oplus T_{\overline{U}\overline{U}} | J = 1) \\
 & =  0
\end{align*}

\item The proof for showing that (\ref{eqn:ach_rate_8}) is satisfied for $\{P_n\}_{n \in \mathbb{N}}$ is similar to showing that (\ref{eqn:ach_rate_7}) is satisfied for $\{P_n\}_{n \in \mathbb{N}}$.

\end{enumerate}


\section{Converse of Theorem~\ref{thm:main}}
\label{apndx:converse_main_result}
The proof of converse is along the lines of the converse arguments in~\cite[Lemma~5]{MishraDPDisit14} (although it does not follow from there). We first argue that following is a general upper bound on $C_{2P}$.
\begin{align*}
 &C_{2P} \leq\\
 &\min\left( \max_{p_X} I(X ; Y | Z),\max_{p_X} I(X ; Z | Y), \max_{p_X} H(X | Y, Z) \right).
\end{align*}
To see that $C_{2P} \leq \max_{p_X} I(X ; Y | Z)$, suppose we run a 2-private data transfer protocol with $U=0$ and $W=1$ (both deterministic). Now $K_0$ is a secret key between Alice and Bob which is secret from Cathy. The bound follows from the fact~\cite{sec-key1993} that the secret key capacity of the broadcast channel $p_{YZ|X}$ with public discussion is upper bounded by $\max_{p_X} I(X ; Y | Z)$. Reversing the roles of Bob and Cathy gives the second term. To prove that $C_{2P} \leq  \max_{p_X} H(X | Y, Z)$, consider running the data transfer protocol with $U=W$, a uniform bit. We may view this as a protocol for two-party OT between Alice and the combination of Bob-Cathy over the channel $p_{YZ|X}$ whose output is $(Y,Z)$. The bound follows from the two-party OT capacity upper bound~\cite{ot2007} of $\max_{p_X} H(X | Y, Z)$. It is easy to evaluate these bound for our binary erasure broadcast channel to obtain the converse:
$\max_{p_X} I(X ; Y | Z) \leq \epsilon_2(1-\epsilon_1)$,\;
$\max_{p_X} I(X ; Z | Y) \leq \epsilon_1(1-\epsilon_2)$,\;
$\max_{p_X} H(X | Y, Z)\leq \epsilon_1\epsilon_2$.

\section{Problem Definition for Databases with $N>2$ Files}
\label{sec:general:setup}
The main difference is that Alice's private database is now made up of $N$ strings $K_0, K_1, \ldots, K_{N-1}$ which are $m$-bit each. Let $\mathbf{K} = (K_0,K_1,\ldots,K_{N-1})$. Bob and Cathy have choice variables $U$ and $W$ respectively which take values in $\{0,1,\ldots,N-1\}$. $\mathbf{K},U,W$ are independent and uniform over their respective alphabets.

Alice's view is now
\[V_A  := (\mathbf{K},X^n,\mathbf{F}),\]
Bob and Cathy's views are given by \eqref{eq:VB}-\eqref{eq:VC}. The privacy conditions \eqref{eqn:ach_rate_3} and \eqref{eqn:ach_rate_7}-\eqref{eqn:ach_rate_8} are replaced by
\begin{align*}
 I(\mathbf{K} \backslash K_U; V_B,V_C| U = W ) & \longrightarrow 0 \\ 
 I(W,\mathbf{K} \backslash K_U ; V_B) & \longrightarrow 0\\ 
 I(U,\mathbf{K} \backslash K_W ; V_C) & \longrightarrow 0,
\end{align*}
where by $\mathbf{S} \backslash \mathbf{T}$ we mean the ordered set $\mathbf{S}$ from which corresponding elements in $\mathbf{T}$ have been removed.   
In addition, we also have a condition to handle the case where $U\neq W$.
\[
 I(\mathbf{K} \backslash (K_U,K_W); V_B,V_C| U \neq W )  \longrightarrow 0. 
\]

\section{Proof of Theorem~\ref{thm:not_so_main}}
\label{apndx:not_so_main_achievability}

To prove the lower bound, we directly extend protocol~\ref{sch:our_2P} to the case where Alice has $N$ strings as follows:

\begin{itemize}
\item Bob now forms $N$ sets $L_0,L_1,\ldots,L_{N-1}$, each of size about $n\min\left( \frac{\epsilon_1}{N-1}, 1 - \epsilon_1 \right)$. The set $L_U$ consists of unerased indices of $Y^n$ and all other sets consist of erased indices of $Y^n$.

\item Cathy confines her attention to $Z^n|_{L_0 \cup L_1 \cup \ldots \cup L_{N-1}}$ and forms her own sets $\tilde{L}_0, \tilde{L}_1, \ldots, \tilde{L}_{N-1}$, each of size about $Nn\min\left( \frac{\epsilon_1}{N-1}, 1 - \epsilon_1 \right) \min\left( \frac{\epsilon_2}{N-1}, 1 - \epsilon_2 \right)$. Only set $\tilde{L}_W$ consists of unerased indices of $Z^n|_{L_0 \cup L_1 \cup \ldots \cup L_{N-1}}$, the other sets contain erased indices of $Z^n|_{L_0 \cup L_1 \cup \ldots \cup L_{N-1}}$.

\item Alice forms the data transfer keys $T_{jj} = X^n|_{L_j \cap \tilde{L}_j}$,  $j = 0,1,\ldots,(N-1)$

\item Alice sends the encrypted strings $M_j = K_j \oplus T_{jj}$, $j = 0,1,\ldots,(N-1)$.

\item Similar to the last two steps of protocol of Section~\ref{sec:achievability}, both Bob and Cathy get extra data transfer rates, using the $2$-party OT protocol \cite{ot2007}, when $\frac{\epsilon_1}{N-1} > 1 - \epsilon_1$ and $\frac{\epsilon_2}{N-1} > 1 - \epsilon_2$.
Alice and Bob use $X^n|_{\tilde{C}}$ (which is completely erased for Cathy) while Alice and Cathy use $X^n|_C$ (which is completely erased for Bob) to obtain this extra data transfer rate $R_{\extra}$. See Appendix~\ref{apndx:sizes_and_extra_rate} for details of all rate calculations.

\end{itemize}
With this modified protocol, achievability of $R_{\lowerb}$ follows along the lines of the proof of Lemma~\ref{lem:achievable_rate}.

The upper bound also immediately follows from the same line of arguments used to establish the converse of Theorem~\ref{thm:main} and a direct extension of the converse of~\cite{ot2007} to 1-out-of-$N$ string OT.


\section{Computing set sizes and data transfer rate expressions}
\label{apndx:sizes_and_extra_rate}
In this section, we will show how the sizes of the different sets that Alice, Bob and Cathy create during the protocol have been calculated. The sizes are given for arbitrary $N$ (number of files). We then derive the expression for the data transfer rate that Bob and Cathy are guaranteed to get in any regime of $\epsilon_1,\epsilon_2$. We finally derive the expression for the extra data transfer rate that Bob and Cathy will get when $\frac{\epsilon_1}{N-1} > 1 - \epsilon_1$ and $\frac{\epsilon_2}{N-1} > 1 - \epsilon_2$.

\subsection{Set Sizes}
\label{apndx:set_sizes}

For ease of notation, let $r_1 = \left(\min\left\{ \frac{\epsilon_1}{N-1}, 1 - \epsilon_1 \right\} - \delta \right)$ and $r_2 = \left(\min\left\{ \frac{\epsilon_2}{N-1}, 1 - \epsilon_2 \right\} - \delta \right)$.

\begin{itemize}
\item $|E| = n(\epsilon_1 - \delta)$
\item $|\overline{E}| = n(1 - \epsilon_1 - \delta)$
\item $ |L_j| = \min\left\{ \frac{|E|}{N-1}, |\overline{E}| \right\} = n r_1$, $j=0,1,\ldots,N-1$
\item $|C| = \left\{  \begin{array}{ll} |E| - (N-1)|\overline{E}|, & \frac{|E|}{N-1} > |\overline{E}| \\ 0, & \frac{|E|}{N-1} \leq |\overline{E}| \end{array} \right.$
\item $|E'| = (|L_0|+|L_1|+\ldots+|L_{N-1}|) \cdot (\epsilon_2 - \delta)$
\item $|\overline{E}'| = (|L_0|+|L_1|+\ldots+|L_{N-1}|) \cdot (1 - \epsilon_2 - \delta)$
\item $|\tilde{L}_j| = \min\left\{ \frac{|E'|}{N-1}, |\overline{E}'| \right\} = N n r_1  r_2$, $j=0,1,\ldots,N-1$
\item $|\tilde{C}| = \left\{  \begin{array}{ll} |E'| - (N-1)|\overline{E}'|, & \frac{|E'|}{N-1} > |\overline{E}'| \\ 0, & \frac{|E'|}{N-1} \leq |\overline{E}'| \end{array} \right.$
\end{itemize}

\subsection{Deriving Data Transfer Rate expressions}
\label{apndx:extra_rate}
The data transfer rate that Bob and Cathy are guaranteed to get in all regimes of $\epsilon_1,\epsilon_2$ is:
\begin{align*}
 R_{\text{guaranteed}} & = \frac{1}{n}|T_{jj}| \\
                                   & = \frac{1}{n} \left( \frac{1}{N}|\tilde{L}_j| \right) \\
                                   & = \frac{1}{n} \left( \frac{1}{N}Nnr_1r_2 \right) \\
                                   & = r_1r_2
\end{align*}

Bob and Cathy get extra data transfer rates when $\frac{\epsilon_1}{N-1} >1 - \epsilon_1$, $\frac{\epsilon_2}{N-1} > 1 - \epsilon_2$. Alice and Bob use $X^n|_{\tilde{C}}$ while Alice and Cathy use $X^n|_C$ for getting this extra rate, using the two-party OT protocol of \cite{ot2007}.

The extra rate Bob \emph{can} get is $|\tilde{C}|\cdot (\frac{1}{N} - \delta)$ while the extra rate Cathy \emph{can} get is $|C|r_2 = |C|(1 - \epsilon_2 - \delta)$. However, since Bob and Cathy can obtain only symmetric rate (see Section~\ref{sec:prob_defn} and Appendix~\ref{sec:general:setup}), the extra rate both Bob and Cathy get is :

\begin{equation*}
R_{\extra} = \min \left\{ |\tilde{C}|\cdot (\frac{1}{N} - \delta),  |C|(1 - \epsilon_2 - \delta) \right\}
\end{equation*}

\end{document}